\documentclass[twocolumn,showpacs,preprintnumbers,amsmath,amssymb,floats]{revtex4}
\usepackage{graphicx}% Include figure files
\usepackage{dcolumn}% Align table columns on decimal point
\usepackage{bm}% bold math

\begin{document}
\title{Helicity order: hidden order parameter in URu$_2$Si$_2$}
\author{C. M. Varma and Lijun Zhu}
\affiliation{Department of Physics, University of California,
  Riverside, CA 92521}

\begin{abstract}
We propose that the ``hidden order parameter" in URu$_2$Si$_2$ is a
helicity order which must arise, if the Pomeranchuk criteria for the
spin-antisymmetric Landau parameters with respect to the stability
of a Fermi liquid state are violated. In a simple model, we
calculate the specific heat, linear and nonlinear magnetic
susceptibilities and the change of transition temperature in a
magnetic field with such an order parameter, and obtain quantitative
agreement with experiments in terms of two parameters extracted from
the data. The peculiar temperature dependence of the NMR linewidth
and the nature of the loss of excitations in the ordered phase seen
by neutron scattering are also explained and experiments are
suggested to directly confirm the proposed order parameter.
\end{abstract}
\maketitle

The ``hidden order" phase in the heavy fermion compound
URu$_2$Si$_2$ below the second order transition at $17.5$K
\cite{Mydosh1985} has remained a puzzle for about 20 years. The
magnitude of the specific heat at the transition is equivalent to
that of ordering of a moment of about 0.5$\mu_B$ per unit cell. No
change in spin-rotational symmetry or lattice translational symmetry
consistent with this specific heat has been discovered. Detailed
neutron diffraction experiments\cite{Broholm1987} reveal a moment of
only about $0.03\mu_B$ per unit cell, which as NMR and $\mu$SR
experiments \cite{Matsuda2001} reveal, is due to the presence of a
second phase. Some very interesting proposals for new types of order
have been made \cite{Santini,Gorkov,Chandra2002}, which have not
been supported by experiments designed to look for them.

Some of the other properties measured at the transition to the
``hidden order" phase and in it are, the linear magnetic
susceptibility which only changes slope at the transition, the
nonlinear magnetic susceptibility which shows a singularity at the
transition similar to that of the specific heat \cite{Ramirez1992},
the change of the transition temperature with an applied magnetic
field \cite{vanDijk1997,Kim2003}, the loss of low energy excitations
observed by neutron scattering for a range of wave vectors
\cite{Broholm1987,Bull2003,Wiebe2004} and in transport
measurements\cite{Palstra1986}, and the NMR relaxation
rate\cite{Bernal2001} which exhibits the extraordinary result that
there is an extra inhomogeneous relaxation rate in the ordered phase
which increases below the transition temperature {\it proportional}
to an order parameter.

We suggest here that the transition is to a state
proposed\cite{cmv2003} as a cure to the spin-antisymmetric
Landau-Pomeranchuk instability(LPI) of the Fermi-liquid. For reasons
which will be clear, we call such states {\it helicity ordered
states}\cite{footnote}. We calculate the thermodynamic properties
near the transition, and account quantitatively for the observed
thermodynamic features and qualitatively for the NMR and the
excitation spectra with parameters extracted from the experiments.
We also suggest experiments which can provide direct evidence for
the proposed phase.

In Landau's Fermi liquid theory\cite{Landau}, the change in the
free-energy due to a a small change of the equilibrium distribution
function $\delta n({\bf k}\sigma)$ is
\begin{eqnarray}
\delta F &=& \sum_{{\bf k}\sigma} \epsilon_{\bf k}^0 \delta n({{\bf
k}\sigma}) \nonumber \\
&&+ {1 \over 2} \sum_{{{\bf k}\sigma}{{\bf k'}\sigma'}} f({{\bf
k}\sigma}, {{\bf k'}\sigma'}) \delta n({{\bf k}\sigma}) \delta
n({{\bf k'}\sigma'}). \label{eq:LFL}
\end{eqnarray}
The interaction functional $f({{\bf k}\sigma},{{\bf k'}\sigma'})$
has spin-symmetric($s$) and spin-antisymmetric($a$) parts:
\begin{equation}
f({{\bf k}\sigma},{{\bf k'}\sigma'})=f^{s}({{\bf k}\sigma,{\bf
k'}\sigma'})+ f^a({\bf k}\sigma, {\bf k'}\sigma').
\label{eq:lfl}
\end{equation}
The coefficients of an expansion of $f^{s,a}$ in terms of the
irreducible representations of the Fermi-surface are the Landau
parameters $F_l^{s,a}$, in terms of which
Pomeranchuk\cite{Pomeranchuk} obtained a set of conditions for the
stability of the Fermi-liquid: $1+(2l+1)^{-1}F_l^{s,a}>0$. Any
violation of these conditions leads to a Landau-Pomeranchuk
instability, which must be cured by a broken symmetry in
corresponding irreducible representation $l$ and spin symmetry $s$
or $a$. For example, the ferromagnetic instability occurs for
$F_0^{a}\to -1$. Spin-symmetric instabilities in finite $l$-channels
have attracted much recent interest\cite{Halboth2000, Fradkin2003,
cmv2003}. A spin-ordered state, which is anisotropic in momentum
space(without change in translational symmetry) is the obvious cure
to the finite $l$ antisymmetric LPIs\cite{cmv2003,Wu2004}. We
develop this idea here; we find that besides the LPI criteria,
additional condition must be satisfied so that  the instability is
of second order.
% and calculate thermodynamic properties and
%correlations near and above the transition.

Consider the model Hamiltonian,
\begin{eqnarray}
 {\cal H}&=&
 \sum_{k\sigma}\epsilon_{k}^0c^{\dag}_{k\sigma}c_{k\sigma} \nonumber
 \\
&+&{1 \over 2} \sum_{{\bf k},{\bf k'};{\bf q}}J_{{\bf k},{\bf
  k'}}({\bf q})
  \left(c^{\dag}_{k+q}\vec{\sigma}c_{k} \right)
  \cdot\left(c^{\dag}_{k'-q}\vec{\sigma}c_{k'} \right),
\label{eq:Hamiltonian}
\end{eqnarray}
where $\epsilon_{k}^0$ is the spectrum of a noninteracting Fermi
gas, $\vec{\sigma}$ Pauli matrices. ${\bf q}$ is the momentum
transfer; of interest is the instability in the forward scattering
limit($q\to 0$). $J_{{\bf k},{\bf k'}}$ is the interaction in
spin-antisymmetric channels, which can be expanded as $J_{{\bf
k},{\bf k'}}(0)=\sum_l J_l P_l(\cos \theta_{{\bf k},{\bf k'}})$,
where $\theta_{{\bf k},{\bf k'}}$ the angle between ${\bf k}$ and
${\bf k}'$, and $P_l(x)$ Legendre polynomials.

In the normal state of a Fermi-liquid, helicity is disordered
since the spin-quantization axes at each ${\bf k}$ can be
independently rotated. The proposed order parameter for the model
has the general  form\cite{cmv2003, Wu2004}:
\begin{equation}
\langle \delta n({\bf k}, \sigma)\rangle =
 \sigma\cdot {\bf D}(\hat{{\bf k}}_f).
\label{eq:op}
\end{equation}
The spin-quantization axis is thereby fixed in relation to the
direction on the Fermi surface\cite{footnote2}.

\begin{figure}
\includegraphics[width=0.35\columnwidth]{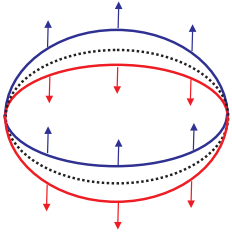}
\caption{Schematic Fermi surface for the two different
spin-directions. The dotted line denotes the Fermi surface of the
paramagnetic phase with vanishing order parameter while the solid
lines illustrate the Fermi surface of the helicity ordered phase.}
\label{fg:fermisurface}
\end{figure}

We need consider only  one specific $l$ channel and write $J_{{\bf
k},{\bf k'}}$ in a separable form $J_l
P_l(\cos\theta_k)P_{l}(\cos\theta_{k'})$. The simplest order
parameter has $D^z(\hat{{\bf k}}_f) \ne 0$, so that the associated
energy parameter is
\begin{equation}
\Delta_l = \langle J_l \sigma^z D^z (\hat{{\bf k}}_f) \rangle =
\langle \sum_k J_l P_l(\cos \theta_k) (n_{k\uparrow} -
n_{k\downarrow}) \rangle.
\label{eq:orderparameter}
\end{equation}
This Ising order parameter is especially useful to discuss the
tetragonal compound URu$_2$Si$_2$, which has a large anisotropy in
the magnetic susceptibility favoring the c-axis. With Eq.
(\ref{eq:orderparameter}) we have a noninteracting model with the
effective spectrum
\begin{equation}
E_{\uparrow, \downarrow}({\bf k})=\epsilon_k^0 \mp \mu_0 H  \pm
\Delta_l P_l(\cos\theta_k),
\label{eq:mf-spec}
\end{equation}
where $H$ is the external magnetic field, and $\mu_0$ the
effective single-electron magnetic moment. Imposing the
requirement of a constant chemical potential, the Fermi surface
for the up and down spins are split as schematically illustration
in Fig.\ref{fg:fermisurface}.

We calculate the free energy following standard methods. The
following results are for $l=1$, but can be easily generalized to
higher-$l$ channels. The free energy can be in general separated
into two parts,
\begin{equation}
{\cal F} = {\cal F}_0(T,H) + {\cal F}_m(T,H,\Delta_1),
\label{eq:free-energy}
\end{equation}
where ${\cal F}_0$ describes a paramagnetic phase($\Delta_1=0$). The
specific heat and  the magnetic susceptibility [$M = \chi_1 H +
(\chi_3/3!) H^3 + \ldots$, $\chi_1$ and $\chi_3$ are the linear,
nonlinear spin susceptibility, respectively] can be easily obtained.
Including terms of $O(H^2)$, $O(T^2)$ and the variation of the
density of states $\rho(\epsilon)$ near the chemical potential to
$O[\rho''(\epsilon_F)]$, we obtain
\begin{subequations}
\begin{eqnarray}
{C_H^0 \over T} &=& {2 \pi^2 \over 3}  k_B^2\rho \left[ 1 +{1\over2}
\left( {\rho'' \over \rho} -{\rho'^2 \over \rho^2}\right)
\left(\mu_0 H\right)^2 \right], \label{eq:cH0}
\\
\chi_1^0 &=&2 \mu_0^2 \rho \left[ 1 +\left( {\rho'' \over
\rho}-{\rho'^2 \over \rho^2}\right) {\pi^2 \over 6} \left(k_B
T\right)^2 \right], \label{eq:chi1-0}
\\
\chi_3^0&=& 3! \mu_0^4 \rho \left[ \left({\rho'' \over
3\rho}-{\rho'^2 \over \rho^2}\right)
%\right.
%\nonumber \\
%&&\left.
+ 3{\rho'^2 \over \rho^2} \left({\rho'' \over \rho}-{\rho'^2 \over
3\rho^2} \right) {\pi ^2 \over 6}(k_BT)^2\right]. \nonumber
\\
 \label{eq:chi3-0}
\end{eqnarray}
\label{eq:thermo-disorder}
\end{subequations}
For noninteracting electrons, these are standard results(see, e.g.,
Ref.\cite{Ashcroft}); we list them here to use them to extract
parameters from the normal state experimental results. For
interacting electrons (in the limit of zero field) they are
multiplied by Landau parameters, $m^*/m$ for the specific heat,
$(m^*/m)/(1+F_0^a)$ for the susceptibility and $(m^*/m)/(1+4F_0^a)$
for the nonlinear susceptibility.

${\cal F}_m(T,H,\Delta_1)$, is the additional contribution to the
free energy for  $\Delta_1 \ne 0$. Expressing it in series of
the order parameter $\Delta_1$ gives,
\begin{eqnarray}
{\cal F}_m&=& {1\over 2} A \Delta_1^2 + {1 \over 4}B \Delta_1^4 ,
\nonumber \\
A&=& {2\rho \over 3} \left( 1+{3\over 2 \rho J_1} \right)
\nonumber \\
&& +{2 \rho \over 3} \left({\rho'' \over \rho}-{\rho'^2 \over
\rho^2} \right) \left[{\pi^2 \over 6}\left(k_B T\right)^2
+{\left(\mu_0 H\right)^2 \over 2}\right]
\nonumber \\
&& + \rho {\rho'^2 \over \rho^2} \left({\rho'' \over \rho}
 -{\rho'^2 \over 3\rho^2} \right)
{\pi^2 \over 6} \left(k_B T\right)^2 \left(\mu_0 H\right)^2,
\nonumber \\
B&=& \rho  \left({\rho'' \over 5 \rho}-{\rho'^2 \over 3 \rho^2}
\right). \label{eq:free-energy-order}
\end{eqnarray}

When $H=0$ and $T=0$, the criterion to have a continuous phase transition
is $A<0$ and $B>0$, {\it i.e.},
\begin{equation}
1+{ 3  \over 2\rho J_1} < 0;
 ~~~~~ {\rho'' \over 5 \rho}- {\rho'^2 \over 3 \rho^2} >0.
\label{eq:pt-condition}
\end{equation}
The first gives $J_1<0$ and $\rho |J_1| > (2l+1)/2$ (here $l=1$),
which is precisely the LPI criterion. The second is an additional
criterion, on the form of the  density of states at the Fermi
surface to have a second order transition. If $B<0$, one must
expand the free energy to terms of order $\Delta_1^6$. In that
case, a first-order phase transition is favored.

At $H=0$, the critical temperature is given by
\begin{equation}
{\pi^2 \over 6} (k_B T_c)^2 = -\left. \left(1+{3\over 2 \rho J_1}
\right) \right/ \left({\rho'' \over \rho}-{\rho'^2 \over \rho^2}
\right). \label{eq:Tc-H0}
\end{equation}
This, together with Eq.(\ref{eq:pt-condition}), requires $\rho'' /
\rho> \rho'^2 / \rho^2$. In presence of a small magnetic field,
$T_c$ varies as
\begin{eqnarray}
T_c^H &\approx& T_c \left[1-(H/H_0)^2 \right], \label{eq:Tc-H}
\\
H_0 &=& {(2/3)^{1/2} { \pi k_B T_c / \mu_0} \over \left[ 1+ {\pi^2
(k_B T_c)^2 \over 2}{\rho'^2 \over \rho^2} h_\rho \right]^{1/2} },
\nonumber
\end{eqnarray}
where $h_\rho\equiv[\rho''/ \rho-\rho'^2 / (3\rho^2)] / (\rho''/
\rho- \rho'^2/ \rho^2)$.

Below $T_c$ we have the non-trivial solution $\Delta_1^2 = - A /B$,
i.e., a helicity-ordered state. The changes in some thermodynamical
quantities at and below $T_c$ from their values for $T>T_c$ are
calculated to be
\begin{subequations}
\begin{eqnarray}
\delta  {c_H}/c_H^0 &=&{\rho \over 81} \pi^4 k_B^4 g_\rho
(3T^2-{T_c}^2), \label{eq:c-m}
\\
\delta \chi_1/\chi_1^0 &=&{2\rho \over 9}{\mu_0^2\over \chi_1^0}
g_\rho {\pi^2 \over 6}k_B^2(T^2-{T_c}^2), \label{eq:chi1-m}
\\
\delta \chi_3/\chi_3^0 &=& {2\rho \over 3} {\mu_0^4\over \chi_3^0}
g_\rho \left[ 1 +  h_\rho{\rho'^2 \over \rho^2} {\pi^2 \over 6} (k_B
T)^2 \right], \label{eq:ch3-m}
\end{eqnarray}
\label{eq:thermo-order}
\end{subequations}
where $g_\rho \equiv (\rho''/\rho-\rho'^2/\rho^2)^2 /
[\rho''/(5\rho) -\rho'^2/(3\rho^2)]$. The specific heat shows
of-course a characteristic mean-field discontinuity at the
transition point; more interesting is the fact that $\chi_3$ also
shows a discontinuity while the linear magnetic susceptibility shows
merely a change of slope. A singularity in the nonlinear magnetic
susceptibility is to be expected for any order parameter $O$ when a
term $|O|^2 H^2$ is allowed in the free energy. Usually the
coefficient of this term is so small that the singularity in
$\chi_3$ is not noticed. What is special about URu$_2$Si$_2$ is that
the dimensionless mean-field jump in $\chi_3$ is similar to the
dimensionless mean-field jump in the specific heat. This and and
several other properties are quantitatively explained below.

We have ignored the Landau parameters in the dimensionless
quantities in Eqs.(\ref{eq:thermo-disorder},
\ref{eq:thermo-order}). The reason is that the change in the
Landau parameters near the transition may be shown following
Leggett \cite{Leggett65}, for the case of transition in superfluid
$^3He$, to be proportional to $(\rho \Delta_1)^2$. To this order,
assumed $\ll1$, they vanish in dimensionless quantities.

\begin{figure}
\includegraphics[width=0.9\columnwidth]{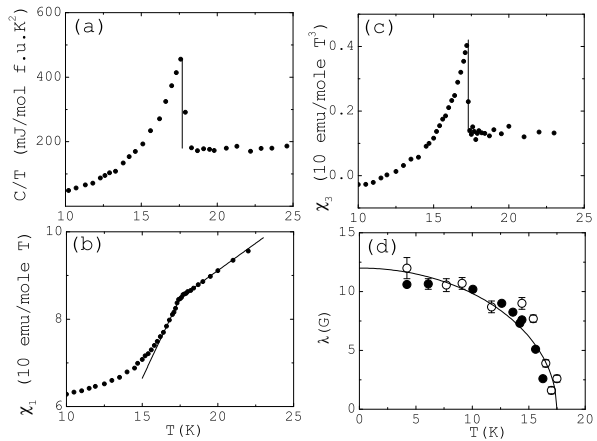}
\caption{The experimental data of direct relevance to the
calculations in this paper. The specific heat data(a) are extracted
from Fig. 1 in Ref.\cite{Mydosh1985}. In the specific heat data a
contribution to $C/T$ proportional to $T^2$, presumably mostly due
to phonons, has been subtracted off.  Data of the linear(b) and
nonlinear(c) magnetic susceptibilities (along the c-axis) are
extracted from Fig. 2 in Ref.\cite{Ramirez1992}. (d) shows the
inhomogeneous linewidth of Si-NMR for magnetic fields in the
c-direction and in the plane, which is extracted from Fig. 4 in
Ref.\cite{Bernal2001}; the solid line is the fitting function
$\lambda=12[1-(T/T_c)^2]^{1/2}$(G).} \label{fg:URu2Si2}
\end{figure}

%\section{URu$_2$Si$_2$}
In the following, we try to fit the experimental data: the specific
heat from Ref.\cite{Mydosh1985}, and the linear and nonlinear
magnetic susceptibilities from Ref.\cite{Ramirez1992}. The data are
shown in Fig.\ref{fg:URu2Si2} for the reader's convenience. From
Eqs. (\ref{eq:thermo-disorder}) and (\ref{eq:thermo-order}), in
addition to the prefactors, $\gamma_0 = {2 \pi^2 \over 3} \rho
k_B^2$, $\chi_0=2\mu_0^2 \rho$, $\tilde{\chi}_0 = 3!\mu_0^4\rho $,
all other quantities can be determined by two additional
dimensionless variables, $C_1 =
(\rho''/\rho-\rho'^2/\rho^2)(\pi^2/6)(k_BT_c)^2$ and
$C_2=(\rho'^2/\rho^2) /(\rho''/\rho)$, where $T_c$ is taken as
$17.5K$.

Consider the linear magnetic susceptibility. It is
continuous with a change in the slope at the transition point,
which is consistent with the  result in Eq.(\ref{eq:chi1-m}). Also
notice that in the normal state it shows a significant linear
temperature dependence besides the constant Pauli term.   This is
indeed required by the theory in order that there be a second
order transition to the helicity ordered phase [see Eq.
(\ref{eq:pt-condition})]. Near $T_c$,
\begin{eqnarray}
\chi_1(T\gtrsim T_c) &=& \chi_0(1+C_1) + 2\chi_0 C_1(T-T_c)/T_c,
\nonumber \\
\delta\chi_1(T\lesssim T_c)&=&\chi_0 {10 \over 9}
 C_1 {1-C_2 \over 1-5C_2/3} (T-T_c)/T_c.
\label{eq:chi1-Tcexpress}
\end{eqnarray}
By fitting them with the experimental data, we can extract
\begin{equation}
C_1 \approx 0.35, ~~~C_2 \approx 0.52,
\label{eq:C1C2}
\end{equation}
(with $\chi_0 \approx 63.0 \text{ emu/mole T}$). We then can
calculate other thermodynamic quantities.  At $T_c$, the
discontinuities of the specific heat coefficient($\gamma=C/T$) and
the nonlinear magnetic susceptibility are calculated from
Eq.(\ref{eq:thermo-order}) to be
\begin{equation}
{\delta \gamma(T_c)/ \gamma_0}= 1.4,~~~
{\delta\chi_3(T_c)/\chi_3(T_c^+)}=2.1, \label{eq:jump-fit}
\end{equation}
while the corresponding experimental quantities from Fig.
\ref{fg:URu2Si2} are approximately 1.5 and 2.4, respectively.  In
finite magnetic field, Eq.(\ref{eq:Tc-H}) predicts that $T_c$
decreases as $(H/H_0)^2$, where $H_0=38.2T$ if we take $\mu_0$ to be
one Bohr magneton. This again, is in agreement with the experiments,
where $H_0$ is estimated as 48.5(1)T in Ref.\cite{vanDijk1997},
while 35.3T in Ref.\cite{Kim2003}. We can also determine the order
parameter ($H=0$)
\begin{equation}
\Delta_1(T)=77\left[1-(T/T_c)^2\right]^{1/2}\text{K}.
\label{eq:op-fit}
\end{equation}

To summarize, we get the qualitatively correct behavior of the
linear susceptibility and extracting two parameters from it, can
explain quantitatively the relative jump in the specific heat and
that of the nonlinear susceptibility as well as the characteristic
field for the suppression of the transition. (The fact that a simple
model for the Fermi surface of the paramagnetic phase gives these
quantities quite well is doubtless due to the fact that we are
comparing  dimensionless quantities.) We do not do well on the
specific heat just below $T_c$; our slope is about a factor of 2
smaller than the straight line fit one might make in
Fig.\ref{fg:URu2Si2}.  Note that the proposed order parameter  has
an Ising symmetry. So, the actual  exponents in the critical regime
of the transition are expected to be that of an Ising model in
3-dimensions; for example the specific heat is expected to show a
lambda shape. While further experiments are required to test this,
the measured specific heat is not inconsistent with such a form (see
Fig. \ref{fg:URu2Si2}). In such a case, a mean-field fit to the data
always gives a slope smaller than the experiment.

The values of $C_1, C_2$ required above imply that $\rho''/\rho
\approx 2 \rho'^2/\rho^2$, i.e., the chemical potential in the
normal phase lies near a local minima of the density of states. This
is consistent with the density of states calculated by
band-structure calculations\cite{Rozing1991}.

Let us next consider the NMR measurements. Si-NMR\cite{Bernal2001}
reveals no change in Knight shift but an increase in the
inhomogeneous linewidth below $T_c$ which within experimental
uncertainty can be fitted to be $\propto [1-(T/T_c)^2]^{1/2}$[see
Fig.\ref{fg:URu2Si2}(d)], i.e. {\it proportional} to an order
parameter. This is quite unusual. We note first that in a
perfectly pure sample, we expect no change in Knight shift or
linewidth. In the presence of impurities, which {\it locally}
break the reflection symmetry about the basal plane, a local
ferromagnetic region forms, as is evident from
Fig.\ref{fg:fermisurface}. The magnitude of the local field then
is proportional to the order parameter but its magnitude as well
as direction are random. This gives no Knight shift but a
linewidth consistent with observations. The magnitude depends on
details of the defect, but a $O(0.5 \mu_B)$ defect, expected from
the magnitude of the order parameter, need be present only in
concentrations of a few parts in a thousand to produce the
observed  linewidth of order $10$ Gauss. The observed linewidth is
almost independent of the direction of the applied magnetic field.
This can be shown to occur for generic distribution of impurities
about the Si-sites\cite{MacLaughlin}.

We intend to calculate the excitation spectra in the future. One
can however see that, given the Fermi-surfaces shown in
Fig.\ref{fg:fermisurface}, a decrease in inelastic scattering
below a certain characteristic energy of the order of $\Delta_1$
is expected for spin-flip particle-hole excitations. This is what
is observed in neutron
scattering\cite{Broholm1987,Bull2003,Wiebe2004}. A quantitative
calculation requires a more realistic model of the normal state
Fermi-surface than used here.  We note that the characteristic
magnitude of this energy scale observed in inelastic neutron
scattering experiments is\cite{Broholm1987} $\sim 70K$, which
compares well with our estimate [see the coefficient in
Eq.(\ref{eq:op-fit})].

Finally, we turn to how the proposed order parameter may be directly
observed. On applying an electric field parallel to the c-axis, a
spin-current would be generated  but no such effect should occur on
applying electric field parallel to the basal plane\cite{Migliori}.
Another direct possibility is through spin-polarized positron
annihilation suggested to us\cite{Mills}.

We wish to acknowledge useful discussions with V. Aji, W. Buyers, D.
MacLaughlin, A. Mills, A. Migliori, J. Mydosh, and A. Ramirez.

\end{document}